\documentclass[aps,prd,twocolumn,showpacs,amsmath,amssymb,floatfix,nofootinbib,eqsecnum]{revtex4-1}

\usepackage{graphicx,color,framed}
\usepackage{hyperref}
\usepackage{times}
\usepackage{enumerate}
\usepackage{lipsum}
\usepackage{slashed}
\usepackage{array}
\usepackage{textcomp}
\usepackage{amsthm}
\usepackage{color}

\hypersetup{
    colorlinks=true, 
    linktoc=all,     
    linkcolor=blue,  
}

\newtheorem{thm}{Theorem}
\newtheorem*{thm*}{Theorem}

\newtheorem*{lemma*}{Lemma}

\def \beq {\begin{equation}}
\def \eeq {\end{equation}}
\def \beqa {\begin{eqnarray}}
\def \eeqa {\end{eqnarray}}
\def \bseq {\begin{subequations}}
\def \eseq {\end{subequations}}
\newcommand \dg {\dagger}

\newcommand \ran {\rangle}
\newcommand \lan {\langle}

\newcommand \ep {\epsilon}

\newcommand \nnb {\nonumber}

\newcommand \td {\tilde}

\newcommand \al {\alpha}

\begin{document}

\title{Stability of topological superconducting qubits with number conservation}

\author{Matthew F. Lapa}
\email[email address: ]{matthew.lapa.physics@gmail.com}
\affiliation{Ocala, FL, 34482, USA}

\author{Michael Levin}
\email[email address: ]{malevin@uchicago.edu}
\affiliation{Kadanoff Center for Theoretical Physics, University of Chicago, Chicago, IL, 60637, USA}


\begin{abstract}

The study of topological superconductivity is largely based on the analysis of simple mean-field models that do not conserve particle number. A major open question in the field is whether the remarkable properties of these mean-field models persist in more realistic models with a conserved total particle number and long-range interactions. For applications to quantum computation, two key properties that one would like to verify in more realistic models are (i) the existence of a set of low-energy states (the qubit states) that are separated from the rest of the spectrum by a finite energy gap, and (ii) an exponentially small (in system size) bound on the splitting of the energies of the qubit states. It is well known that these properties hold for mean-field models, but so far only property (i) has been verified in a number-conserving model. In this work we fill this gap by rigorously establishing both properties (i) and (ii) for a number-conserving toy model of two topological superconducting wires coupled to a single bulk superconductor. Our result holds in a broad region of the parameter space of this model, suggesting that properties (i) and (ii) are robust properties of number-conserving models, and not just artifacts of the mean-field approximation.

\end{abstract}


\maketitle

\section{Introduction}

Although over 20 years have passed since the original work on Majorana fermions (MFs) in condensed
matter systems~\cite{kitaev,read-green}, research on topological superconductors (TSCs)
shows no signs of slowing down. One of the main reasons
for this sustained activity is the possibility that these systems could be used to perform fault-tolerant
quantum computation~\cite{kitaev}. A great deal of theoretical and experimental work has
been done to realize TSC phases in the laboratory and to devise schemes for performing protected
quantum computing operations using MFs~\cite{PhysRevLett.105.077001,
PhysRevLett.105.177002,PhysRevB.88.020407,
mourik2012signatures,das2012zero,rokhinson2012fractional,
deng2012anomalous,PhysRevLett.110.126406,
PhysRevB.87.241401,yazdani2014observation,albrecht2016exponential,alicea-milestones}. 
Despite these efforts, there is still no broad consensus on whether MFs have indeed been
observed in the laboratory~\cite{zb1,zb2,ji-wen2018,sau2018,absence2020,yu2021non,frolov2021,
leggett-lin2022} (see Ref.~\onlinecite{frolov2021} for a review and critique of the current 
experimental situation). 

In the last decade several researchers raised an important theoretical concern about these 
efforts~\cite{Leggett1,Leggett2,ortiz2,lin-leggett1,lin-leggett2,leggett-lin2022}.
Almost all of the calculations involved in the work on TSCs rely on the \emph{mean-field} 
(or Bogoliubov-de Gennes) approach to superconductivity, an approach that violates particle
number conservation symmetry. 
In particular, Leggett noted that although the mean-field approach
accurately captures the bulk thermodynamic properties 
of many superconductors, the mean-field
eigenstates may not represent the true eigenstates of a superconductor accurately enough
for quantum computing applications (which depend on the detailed properties of individual quantum 
states).

In our own work~\cite{LL} on this subject we raised an additional concern.
While most work on TSCs relies on mean-field
models with short-range interactions, real charged superconductors are described by 
number-conserving Hamiltonians with \emph{long-range} interactions.\footnote{All gapped superconductors must contain long-range interactions, irrespective of the physical realization, as can be seen from the Goldstone theorem of Ref.~\onlinecite{hastings1}.} 
Because of this, there is no guarantee that the usual mean-field models can accurately capture the 
topological properties of real charged superconductors.

There are two key properties of the mean-field TSC models that are
important for quantum computing applications and that one would like to verify in a number-conserving
model. These are (i) the existence of a set of low-energy states (the qubit states) that are separated from 
the rest of the spectrum by a finite energy gap, and (ii) an exponentially small (in system size) bound on 
the splitting of the energies of the qubit states.

In the mean-field models these properties
are closely connected to the existence of unpaired MFs at specific locations in the
system (e.g., the two ends of a 1D wire or the cores of quantum vortices in a 2D system).
The presence of these MFs also leads to unusual long-range correlations between their
locations (typically the MFs are separated by a distance on the order of the system 
size)~\cite{pfeuty,SS1,SS2,fu2010,vijay-fu,reslen2018end}. 
Therefore we can add to the above two properties a third interesting property: (iii) the
existence of long-range ``Majorana-like'' correlations.

In Ref.~\cite{LL} we investigated whether these properties of mean-field TSC models carry over to the number-conserving setting. Specifically, we focused on a number-conserving toy model for a 1D proximity-induced topological superconductor. This model consisted of a fermionic wire coupled to a number and phase degree of freedom representing a bulk s-wave superconductor.
For this toy model, we proved that 
properties (i) and (iii) hold for a wide range of parameter values -- i.e.~without any fine-tuning of parameters.
However, the methods we used
in Ref.~\onlinecite{LL} were not strong enough to address property (ii), and so the question of the
energy splitting of the qubit states remained open.\footnote{The methods from 
Ref.~\onlinecite{LL} (when applied to the two-wire model in this paper) are only strong enough to 
prove a \emph{power-law} bound on the splitting. Specifically, they can be used to obtain a
bound that scales as $L^{-1}$, where $L$ is the length of the topological superconducting wires.} 

In this paper we fill this gap in the literature by rigorously establishing property (ii) for a number-conserving toy model for a 1D TSC. In other words, we prove that the energy splitting between the two qubit states in our toy model is exponentially small in the system size. As in Ref.~\cite{LL}, our result holds in an open region of the model's parameter 
space (i.e., it does not require fine-tuning). This robust, exponentially small splitting is important because it means that the TSC qubit will take an exponentially long time to decohere, even in a number conserving system.

The main idea of our proof is to use a stability result for quantum spin systems with long-range 
interactions that we proved in Ref.~\onlinecite{LL-stability}. After some manipulations, we show
that our TSC model can be mapped onto the type of spin system studied in 
Ref.~\onlinecite{LL-stability}, and then we invoke the stability result from that paper to complete 
the proof of the exponential splitting property.
As an interesting side note, this approach also provides an alternative proof of property (i) in our
TSC model.

The model that we study in this paper is a two-wire version of our model from Ref.~\cite{LL}.
It is known that two wires are needed for any qubit setup
based on a 1D TSC --- in the mean-field or number-conserving settings --- for the following reason. 
In the mean-field setting, if we want to use two eigenstates of a system to form a qubit, then those 
two states need to have the same fermion parity. This is because of a 
\emph{superselection rule} that forbids the existence of quantum superpositions of states with opposite
fermion parity. Therefore, to construct a qubit using the low-energy states of a 1D TSC, we actually
need two fermionic wires (the low-energy states of one wire have opposite fermion 
parity)~\cite{alicea-milestones}. In that case there are four 
low-energy states, and two of these states with the same 
parity can be used to form a qubit. In the number-conserving setting the situation is almost the same
except now the superselection rule forbids superpositions of states with different fermion 
number.\footnote{Superselection rules in this context were discussed in detail in 
Refs.~\onlinecite{ortiz1,ortiz2}.}
Again, two wires are necessary to obtain two low-energy states with the same fermion number.

Topological superconductivity with number conservation has been studied previously using several
different methods. One approach uses bosonization to study low-energy field theory 
models in 1D~\cite{cheng-tu,fidkowski2011,Sau2011,tsvelik2011zero,cheng-lutchyn, 
KSH,knapp2019number,RuhmanBergAltman}. This approach delivers general results but requires various
approximations. In addition, because these models are one-dimensional and have only local interactions, they
only feature quasi-long range order, as opposed to the true superconducting order present in our model. 
A second approach is based on exactly-solvable 
models~\cite{PhysRevB.82.224510,ortiz1,ortiz2,iemini2015localized,lang2015topological,PhysRevB.96.115110},
including models in 1D and 2D. This approach yields rigorous results but requires
fine-tuning of the parameters in each model. Finally, several works have studied these
systems using numerical methods or other kinds of approximate analytical 
arguments~\cite{Kraus2013,burnell2018,RA,lin-leggett1,lin-leggett2,cui2019,leggett-lin2022}.
The key differences between our work and these previous results are (i) we have been
able to obtain rigorous results on a concrete model, and (ii) our results are robust in the
sense that they do not require fine-tuning of the parameters in our model.\footnote{In a separate
line of work~\cite{Lapa2020}, one of us obtained rigorous results on topological invariants for TSCs 
in the number-conserving setting. Those
results are also robust in that they hold for any number-conserving model that can be adiabatically 
connected to a gapped pairing model of a TSC.}  

This paper is organized as follows. In Sec.~\ref{sec:model-main-results} we introduce our model and
state our main stability results. In Sec.~\ref{sec:proof} we present the proof of our main result. The 
bulk of this section is dedicated to explaining the mapping from our TSC model to a spin model of the
kind that we studied in Ref.~\onlinecite{LL-stability}. In Sec.~\ref{sec:discussion-conclusion} we
present our conclusions. Finally, in Appendix~\ref{app:changes-to-proof} we discuss some of the minor 
technical changes that need to be made to our proof from Ref.~\onlinecite{LL-stability} in order to apply 
those results to the two-wire TSC model in this paper.

\section{Model and main results}
\label{sec:model-main-results}

\subsection{A number-conserving model of two topological superconducting wires}

\begin{figure}[t]
  \centering
    \includegraphics[width= .45\textwidth]{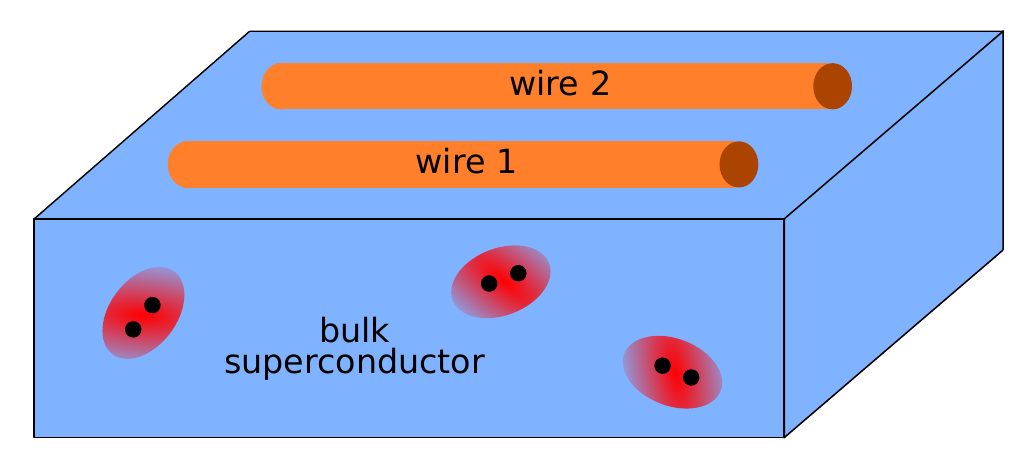} 
\caption{The physical picture for the TSC model studied in this paper: two fermionic wires are
placed on top of a bulk s-wave superconductor. The wires interact with the bulk superconductor by 
exchanging pairs of fermions, and the total number of fermions in the wires plus the superconductor
is always conserved.}
\label{fig:TSC}
\end{figure}

The model that we study in this paper is a two-wire version of the model that we studied in 
Ref.~\onlinecite{LL}. The degrees of freedom in this model consist of (i) spinless fermions living on 
two quantum wires, and (ii) a number and phase degree of freedom that serves as a toy model for a bulk
s-wave superconductor (SC). Both quantum wires have open boundary conditions and should be
thought of as finite segments of wire deposited on top of a bulk SC, as shown in Fig.~\ref{fig:TSC}.

We label the two wires by $\al\in\{1,2\}$ and we label the sites on each wire
by $j\in\{1,\dots,L\}$, where $L$ is the length of each wire. The operator that
annihilates a spinless fermion on site $j$ and wire $\al$ is 
$c_{j,\al}$, and these operators obey the standard anticommutation relations: 
$\{c_{j,\al},c_{j',\beta}\}=0$ and $\{c_{j,\al},c^{\dg}_{j',\beta}\}= \delta_{jj'}\delta_{\al\beta}$. 
We define the number operators $n_{j,\al}=c^{\dg}_{j,\al}c_{j,\al}$ in the usual way. 
We also define the number operator $\mathcal{N}_{w,\al}$ for wire $\al$ as
$\mathcal{N}_{w,\al}= \sum_{j=1}^L n_{j,\al}$. The number operator $\mathcal{N}_w$ for both wires together 
is then given by $\mathcal{N}_w= \sum_{\al=1}^2\mathcal{N}_{w,\al}$.

The number and phase degree of freedom that represents the bulk SC is comprised of the operators 
$\hat{n}$ and $\hat{\phi}$.
We use a hat for these operators in order to distinguish the phase operator $\hat{\phi}$ from classical
phases $\phi$ that appear in various parts of our analysis. The operator $\hat{n}$ is defined to have
integer eigenvalues, and $\hat{n}$ and $\hat{\phi}$ obey the usual commutation relations
$[\hat{\phi}, \hat{n}] = i$.
The Hilbert space $\mathcal{H}_{\text{SC}}$ of the number and phase degree of freedom is spanned by
the states $|p\ran$, where $p\in\mathbb{Z}$ and we have
$\hat{n}|p\ran=p|p\ran$ and $e^{\pm i\hat{\phi}}|p\ran=|p\pm 1\ran$. The total Hilbert space of
our model is then the tensor product $\mathcal{F}\otimes\mathcal{H}_{\text{SC}}$, where $\mathcal{F}$
is the Fock space for the fermions $c_{j,\al}$ on the two wires.

The operator $\hat{n}$ counts the number of Cooper pairs in the bulk SC, and $e^{i\hat{\phi}}$
adds a Cooper pair to the bulk SC (similarly, $e^{-i\hat{\phi}}$ removes a Cooper pair). Therefore, the
total number operator for our entire system is 
\beq
	\mathcal{N}= \mathcal{N}_w + 2\hat{n}\ ,
\eeq
where $\mathcal{N}_w$ is the number operator for the two wires, and
the factor of two is present because each Cooper pair has twice the charge of a single fermion.
The model that we study in this paper conserves the total particle number of the two wires 
\emph{plus} the bulk SC, and so our model Hamiltonian will commute with $\mathcal{N}$.

The Hamiltonian for our model takes the general form
\beq
	H=H_0+V\ .
\eeq
The first term $H_0$ resembles a standard mean-field Hamiltonian~\cite{kitaev} for two decoupled $p$-wave wires with open boundary conditions, but with the important difference that the wires are coupled to the bulk SC in such a way that the total particle number $\mathcal{N}$ is conserved. Specifically, $H_0$ is given by
\begin{align}
	H_0 =& -\sum_{\al=1}^2\frac{t_{\al}}{2}\sum_{j=1}^{L-1}(c^{\dg}_{j,\al} c_{j+1,\al}+\text{h.c.}) \nnb \\
	-&	\sum_{\al=1}^2\frac{\Delta_{\al}}{2}\sum_{j=1}^{L-1}(c_{j,\al} c_{j+1,\al}e^{i\hat{\phi}}+\text{h.c.}) \nnb \\
	-& \sum_{\al=1}^2 \mu_{\al} \sum_{j=1}^L\left(n_{j,\al} -\frac{1}{2}\right)\ .
\end{align}
The first line contains nearest-neighbor hopping terms for the fermions in both wires, with
hopping energies $t_{\al}$ that we take to be real and positive. The second line contains nearest-neighbor
pairing terms in each wire, with pairing energies $\Delta_{\al}$ that we assume to be real numbers. Notice
that, unlike a normal mean-field Hamiltonian, the product $c_{j,\al} c_{j+1,\al}$ appears together 
with the phase \emph{operator} $e^{i\hat{\phi}}$, and so the pairing term in $H_0$ 
commutes with the total particle number operator 
$\mathcal{N}$. The physical interpretation of the term $c_{j,\al} c_{j+1,\al}e^{i\hat{\phi}}$ is that
two fermions leave wire $\al$ and enter the bulk SC as a Cooper pair. 
Finally, the third line in $H_0$ contains a chemical potential term for each wire, with chemical
potentials $\mu_{\al}$ that are real numbers.

The second term $V$ is a charging energy term that gives an energy cost for transferring charge from the 
bulk SC to the wires or vice-versa. This term takes the form\footnote{The reader may wonder
why $V$ does not include a quadratic term for the charge of the bulk SC, or 
cross terms that couple the charge on each wire to that of the bulk SC. The reason for this is that, when 
we restrict to a sector 
of fixed total particle number $\mathcal{N}$, those terms can be rewritten in the same form as the terms 
that are already included in $V$.}
\begin{align}
	V &= \frac{1}{2}\sum_{\al=1}^2 E_{c,\al}(\mathcal{N}_{w,\al} - N_{\al,0})^2 \nnb \\
	 &+ E_c'(\mathcal{N}_{w,1} - N_{1,0})(\mathcal{N}_{w,2} - N_{2,0})\ ,
\end{align}
where the parameters appearing in this expression are as follows. First, $N_{\al,0}$ is the electrostatically preferred number of electrons in wire $\alpha$, which is presumably equal to the number of electrons required to make wire $\al$ charge neutral.
Next, $E_{c,\al}$ and $E_c'$ 
are (inverse) capacitance coefficients that describe the electrostatic energy cost for deviating from charge neutrality. On physical
grounds we have $E_{c,\al},E_c' \geq 0$, and in order to make $V$ positive definite we also require that 
\beq
	E_{c,1}E_{c,2} \geq (E_c')^2\ .
\eeq

As in Ref.~\onlinecite{LL}, we parametrize $E_{c,\al}$ and $E_c'$ as
\beq
	E_{c,\al} = \frac{\mathcal{E}_{c,\al}}{L}\ \ ,\ \ E_c' = \frac{\mathcal{E}_c'}{L}\ ,
\eeq
where $\mathcal{E}_{c,\al}$ and $\mathcal{E}_c'$ are rescaled charging energies that we hold fixed in the 
thermodynamic limit $L\to\infty$. The general form of the charging energy term $V$, and the scaling of
$E_{c,\al}$ and $E_c'$ with $L$, follow from electrostatics considerations and capture the influence of the
Coulomb interaction on processes in which charge is exchanged between the wires and the bulk SC. In 
particular, $V$ is the physically correct form of the charging energy when the linear dimensions of the bulk 
SC are comparable to the length $L$ of the wires, and when the length of the wires is much larger than their
thickness and their separation from the SC.

One key feature of our model is that it does not include processes in which a \emph{single} fermion enters or leaves the SC; the SC can only accommodate pairs of fermions. Our model also does not include processes where a single fermion tunnels from one wire to the other.

The physical picture that motivates these properties is as follows. The reason we neglect single fermion excitations within the SC is that we assume that the energy gap to such excitations is the largest energy scale in the system, and we think of our model as a low energy effective theory describing physics below this energy scale. Likewise, the reason we neglect single fermion tunneling between the wires is that we assume that the wires are sufficiently far apart from one another that such tunneling processes are negligible.

\subsection{Main stability results}
We now state our main stability result for our two-wire TSC model. Our result concerns the properties of the model
in a sector of fixed total particle number equal to $N$ (the ``$N$-particle sector''). 
In other words, we work in the sector of the total Hilbert
space in which $\mathcal{N}=N$ (and $N$ can be any positive integer). 
Each sector of fixed total particle number can be further
divided into two subsectors, with each subsector containing states with a fixed eigenvalue 
($\pm 1$) of $\mathcal{P}_1 := (-1)^{\mathcal{N}_{w,1}}$, the fermion parity operator for the first wire.
Since $H$ also commutes with $\mathcal{P}_1$, we can separately diagonalize $H$ in each such
subsector of the $N$-particle sector.

Let $|\psi^{(N)}_{\pm}\ran$ be the ground state of $H$ in the subsector of the $N$-particle sector 
with $\mathcal{P}_1=\pm 1$, and let $E^{(N)}_{\pm}$ be the corresponding ground state energy.
Our main goal is to determine the stability of a qubit built out of the two states $|\psi^{(N)}_{+}\ran$ and $|\psi^{(N)}_{-}\ran$. 

More precisely, we wish to determine whether the qubit states $|\psi^{(N)}_{\pm}\ran$ have the following two properties for a finite range of parameters (i.e.~without fine tuning).
First, $H$ should have a finite
energy gap above $E^{(N)}_{\pm}$ in the subsector of the $N$-particle sector with 
$\mathcal{P}_1=\pm 1$. This prevents each qubit state from mixing (e.g., due to thermal fluctuations
at low but non-zero temperatures) with other states in the
same subsector of the $N$-particle sector. Second, the energy splitting 
$|E^{(N)}_{+}-E^{(N)}_{-}|$ between the two qubit states should be exponentially small 
in the system size $L$. This property is known to hold in the mean-field p-wave wire model, and
it guarantees that the qubit will take a very long time to decohere. 

The main result of this paper is a rigorous proof of the second property: we derive an exponential bound on
the energy splitting $|E^{(N)}_{+}-E^{(N)}_{-}|$ between the two qubit states, and we show that this bound holds for a wide range of parameter values. Conveniently, the methods we use to prove this bound on the splitting also provide a proof of the first property -- i.e.~the existence of a finite energy gap in each of the two subsectors of the $N$-particle sector. We note that the existence of a finite energy gap can also be established by a straightforward generalization of the arguments from Ref.~\onlinecite{LL}.

We need to define a few more parameters before we can state our main result. The first of these
are a set of modified chemical potentials $\td{\mu}_{\al}$ for the two wires:
\begin{subequations}
\label{eq:mu-mod}
\begin{align}
	\td{\mu}_1 &= \mu_1 +  \frac{\mathcal{E}_{c,1}}{L}\left(N_{1,0}-\frac{L}{2}\right) +  \frac{\mathcal{E}_c'}{L}\left(N_{2,0}-\frac{L}{2}\right) \\
	\td{\mu}_2 &= \mu_2 +  \frac{\mathcal{E}_{c,2}}{L}\left(N_{2,0}-\frac{L}{2}\right) + \frac{\mathcal{E}_c'}{L}\left(N_{1,0}-\frac{L}{2}\right)\ .
\end{align}
\end{subequations}
We will see why the $\td{\mu}_1$ and $\td{\mu}_2$ parameters are natural in Sec.~\ref{sec:proof} when we map our Hamiltonian onto a spin model, but the basic idea is that they come from expanding the charging energy terms around half-filling.

Finally, we introduce one more parameter $\lambda$, which is defined to be the largest (in magnitude) of all the charging energies and modified chemical potentials $\td{\mu}_{\al}$:
\begin{align}
	\lambda &= \max(|\mathcal{E}_{c,1}|, |\mathcal{E}_{c,2}|, |\mathcal{E}_c'|, |\td{\mu}_1|, |\td{\mu}_2|)\ . 
\end{align}

We will present our result for a particular choice of $t_\al, \Delta_\al$, namely $t_{\al}=\Delta_{\al}$. 
The reason for this choice is as follows: for a single copy of Kitaev's p-wave wire model, it is known that the topological superconductor phase corresponds
to the region in parameter space in which $|\mu|<t$ and $\Delta\neq 0$. In particular, for 
any non-zero value of $\Delta$, the transition from the topological to the trivial superconducting phase 
always occurs at $|\mu|=t$. Since the location of the transition does not depend on the precise value of 
$\Delta$, we are free to pick whatever $\Delta$ is most convenient. Here we choose to study our model at the point $t_{\al}=\Delta_{\al}$ as this makes our
stability analysis of the model more straightforward and transparent. If we wanted to, we could 
prove similar results for small deviations from the point $t_{\al}=\Delta_{\al}$ by treating the 
deviation as an additional short-range perturbation of the unperturbed Hamiltonian with 
$t_{\al}=\Delta_{\al}$.

We are now ready to state our main result. 

\begin{thm}
Consider the two wire Hamiltonian $H$ at the special point $t_{\al}=\Delta_{\al}$. 
There exists a $L$-independent constant $\lambda_0 > 0$
such that, if $\lambda < \lambda_0$, then (1) $H$
has a unique ground state and a finite energy gap in each subsector of the $N$-particle sector with
fixed $\mathcal{P}_1$ eigenvalue, and (2) the energy splitting $|E^{(N)}_{+}-E^{(N)}_{-}|$
between the ground states in the two subsectors of the $N$-particle sector satisfies the exponential bound
\beq
	|E^{(N)}_{+}-E^{(N)}_{-}| \leq c_1 L e^{-c_2 L}\ ,
\eeq
where $c_1$ and $c_2$ are fixed positive constants that depend on 
$\Delta_{\al}$, $\td{\mu}_{\al}$, $\mathcal{E}_{c,\al}$, and $\mathcal{E}_c'$, but not on $L$. 
\label{thm:main}
\end{thm}

This theorem shows that, for small enough $|\td{\mu}_{\al}|$, $|\mathcal{E}_{c,\al}|$, 
and $|\mathcal{E}_c'|$, our
number-conserving TSC model has \emph{both} of the stability properties (i) and (ii) 
that we discussed above. (As a side note, it is worth pointing out that the theorem holds for any signs of the charging energy terms,
though, in our setting, the model only makes physical sense when the charging energy terms are positive). 

\section{Proof of the main result}
\label{sec:proof}

In this section we present the proof of Theorem~\ref{thm:main}. The core of the proof is the
general stability result for spin Hamiltonians with long-range interactions established
in Ref.~\onlinecite{LL-stability}. To apply that result to our model, we first need to map our
model to a suitable spin Hamiltonian. We do that in two steps. In the first step we show that, in the 
$N$-particle sector, our model is equivalent to a purely fermionic model without number 
conservation and restricted to the sector of 
fixed fermion parity equal to $(-1)^N$. In the second step, we convert this purely fermionic model to a spin model 
using a Jordan-Wigner transformation. 

\subsection{Relation to a fermionic model}

We begin by mapping our full model to a purely fermionic model. Our
discussion in this subsection closely follows Sec.~III of the Supplemental Material from 
Ref.~\onlinecite{LL}. 

We start by considering the mean-field Hamiltonian 
$H_{\text{MF}}(\phi)$ that is obtained by replacing the operator $e^{i\hat{\phi}}$ in $H_0$ with the 
classical phase $e^{i\phi}$,
\begin{align}
	H_{\text{MF}}(\phi) =& -\sum_{\al=1}^2\frac{t_{\al}}{2}\sum_{j=1}^{L-1}(c^{\dg}_{j,\al} c_{j+1,\al}+\text{h.c.}) \nnb \\
	-&	\sum_{\al=1}^2\frac{\Delta_{\al}}{2}\sum_{j=1}^{L-1}(c_{j,\al} c_{j+1,\al}e^{i\phi}+\text{h.c.}) \nnb \\
	-& \sum_{\al=1}^2 \mu_{\al} \sum_{j=1}^L\left(n_{j,\al} -\frac{1}{2}\right)\ .
\end{align}
The Hamiltonian $H_{\text{MF}}(\phi)$ acts on only on the Fock space $\mathcal{F}$ for the fermions 
$c_{j,\al}$. The number and phase degree of freedom is not involved in $H_{\text{MF}}(\phi)$. 

Let $|\chi_a(\phi)\ran$, with $a\in\{0,1,2,\dots\}$, be the eigenstates of $H_{\text{MF}}(\phi)$
in the sector of the Fock space with fermion parity $(-1)^N$, and let $\ep_a$ be the energies
of these states: $H_{\text{MF}}(\phi)|\chi_a(\phi)\ran= \ep_a |\chi_a(\phi)\ran$. The 
energies $\ep_a$ are independent of $\phi$ due to the fact that $H_{\text{MF}}(\phi)$ is
related to $H_{\text{MF}}(0)$ by a unitary transformation: 
$H_{\text{MF}}(\phi)= e^{-i\frac{\phi}{2}\mathcal{N}_w}H_{\text{MF}}(0) e^{i\frac{\phi}{2}\mathcal{N}_w}$.
This transformation also implies that the eigenstates of $H_{\text{MF}}(\phi)$ are related to the 
eigenstates of $H_{\text{MF}}(0)$ as 
$|\chi_a(\phi)\ran= e^{-i\frac{\phi}{2}\mathcal{N}_w}|\chi_a(0)\ran$.

It turns out that there is a one-to-one correspondence between the mean-field eigenstates $|\chi_a(\phi)\ran$ and the eigenstates of the number-conserving Hamiltonian $H_0$ in a fixed $N$-particle sector. To explain this correspondence, let $|p\ran$ be the eigenstates of the Cooper pair number operator $\hat{n}$, and let $|\phi\ran$ be the dual set of eigenstates of $e^{i\hat{\phi}}$ that are related to $|p\ran$ by
\beq
	\lan \phi |p\ran= \frac{1}{\sqrt{2\pi}}e^{i p\phi}\ .
\eeq
In this notation, the eigenstates of $H_0$ in the $N$-particle sector take the form\footnote{Our notation 
here highlights the fact that the state $|\chi^{(N)}_a\ran$ lives in the tensor product 
$\mathcal{F}\otimes\mathcal{H}_{\text{SC}}$ of the fermionic 
Fock space $\mathcal{F}$ and the Hilbert space $\mathcal{H}_{\text{SC}}$ for the number and phase degree of 
freedom.}
\beq
	|\chi^{(N)}_a\ran= \frac{1}{\sqrt{2\pi}}\int_0^{2\pi}d\phi\ e^{i\frac{N}{2}\phi} |\chi_a(\phi)\ran\otimes|\phi\ran \ . \label{eq:number-conserving-eigenstates}
\eeq
It is not hard to check that these states satisfy $H_0|\chi^{(N)}_a\ran=\ep_a |\chi^{(N)}_a\ran$ and 
$\mathcal{N}|\chi^{(N)}_a\ran=N|\chi^{(N)}_a\ran$ (see e.g. Sec. III of the Supplemental Material from Ref.~\onlinecite{LL} for details). In addition, these states have
the following important property. 
Let $\mathcal{O}$ be any operator formed from the
fermionic creation and annihilation operators $c_{j,\al}$ and $c^{\dg}_{j,\al}$ that also commutes
with the particle number operator for the wires, $[\mathcal{O},\mathcal{N}_w]=0$. Then
for any two states $|\chi^{(N)}_a\ran$ and $|\chi^{(N)}_b\ran$, we have
the identity
\beq
	\lan\chi^{(N)}_a| \mathcal{O} |\chi^{(N)}_b\ran = \lan\chi_a(0)| \mathcal{O} |\chi_b(0)\ran\ . 
	\label{eq:O-matrix-elements}
\eeq

To prove Eq.~\eqref{eq:O-matrix-elements}, we simply evaluate the matrix element 
$\lan\chi^{(N)}_a| \mathcal{O} |\chi^{(N)}_b\ran$ using our formula 
Eq.~\eqref{eq:number-conserving-eigenstates} for
$|\chi^{(N)}_a\ran$, the orthogonality relation
$\lan\phi|\phi'\ran=\delta(\phi-\phi')$, the fact that 
$|\chi_a(\phi)\ran= e^{-i\frac{\phi}{2}\mathcal{N}_w}|\chi_a(0)\ran$, and our assumption that 
$[\mathcal{O},\mathcal{N}_w]=0$. We have
\begin{widetext}
\beqa
	\lan\chi^{(N)}_a| \mathcal{O} |\chi^{(N)}_b\ran &=& \frac{1}{2\pi}\int_0^{2\pi} d\phi\int_0^{2\pi} d\phi'\  e^{-i\frac{N}{2}\phi}e^{i\frac{N}{2}\phi'} \lan\chi_a(\phi)| \mathcal{O} |\chi_b(\phi')\ran \lan\phi|\phi'\ran \nnb \\
	&=& \frac{1}{2\pi}\int_0^{2\pi} d\phi\int_0^{2\pi} d\phi'\  e^{-i\frac{N}{2}\phi}e^{i\frac{N}{2}\phi'} \lan\chi_a(\phi)| \mathcal{O} |\chi_b(\phi')\ran \delta(\phi-\phi') \nnb \\
	&=& \frac{1}{2\pi}\int_0^{2\pi} d\phi\   \lan\chi_a(\phi)| \mathcal{O} |\chi_b(\phi)\ran \nnb \\
	&=& \frac{1}{2\pi}\int_0^{2\pi} d\phi\   \lan\chi_a(0)|e^{i\frac{\phi}{2}\mathcal{N}_w} \mathcal{O} e^{-i\frac{\phi}{2}\mathcal{N}_w}|\chi_b(0)\ran \nnb \\
	&=& \frac{1}{2\pi}\int_0^{2\pi} d\phi\   \lan\chi_a(0)| \mathcal{O} |\chi_b(0)\ran \nnb \\
	&=& \lan\chi_a(0)| \mathcal{O} |\chi_b(0)\ran\ .
\eeqa
\end{widetext}

With these results, we can now map $H$ onto a purely fermionic Hamiltonian. To do this, we consider
the matrix elements of $H$ in the $|\chi^{(N)}_a\ran$ basis. First, we clearly have
\beq
	\lan\chi^{(N)}_a| H_0 |\chi^{(N)}_b\ran= \ep_a \delta_{ab}= \lan\chi_a(0)| H_{\text{MF}}(0) |\chi_b(0)\ran \ ,
\label{H0id}
\eeq
which follows directly from the construction of the states $|\chi^{(N)}_a\ran$ from the states
$|\chi_a(\phi)\ran$.
Next, for the charging energy term we also find that
\beq
		\lan\chi^{(N)}_a| V |\chi^{(N)}_b\ran = \lan\chi_a(0)| V |\chi_b(0)\ran\ ,
\label{Vid}
\eeq
which follows immediately from Eq.~\eqref{eq:O-matrix-elements} and the fact that $V$ depends only on the
number operators $\mathcal{N}_{w,\al}$ for the wires. 
Adding together (\ref{H0id}) and (\ref{Vid}), we conclude that the matrix elements 
$\lan\chi^{(N)}_a| H |\chi^{(N)}_b\ran$ of our Hamiltonian in the $N$-particle sector are
equal to the matrix elements $\lan\chi_a(0)| H_{\text{F}} |\chi_b(0)\ran$ of a number-non-conserving
fermionic Hamiltonian $H_{\text{F}}$, defined by 
\beq
	H_{\text{F}}= H_{\text{MF}}(0) + V \ , \label{eq:H_F}
\eeq
in the sector with fermion parity equal to $(-1)^N$.

This completes the mapping of our model (within a sector of fixed
total particle number) to a purely fermionic model without number conservation (and restricted to a 
sector of fixed fermion parity). 

\subsection{Mapping to a spin model via Jordan-Wigner transformation}

We now move on to the second step in the mapping to a spin Hamiltonian, which is to use
a Jordan-Wigner (JW) transformation to map the fermionic Hamiltonian $H_{\text{F}}$
from Eq.~\eqref{eq:H_F} into a spin Hamiltonian. Since the JW transformation is a standard method,
we give only a brief description of the most important points.

First, as we discussed in Sec.~\ref{sec:model-main-results}, to simplify our analysis we work at the 
point in parameter space where $t_{\al}=\Delta_{\al}$. 
At this point it is helpful to first rewrite $H_{\text{F}}$ in terms of
Majorana fermion operators before doing the JW transformation. For each wire $\al$ we define 
two Majorana fermion operators $a_{j,\al}$ and $b_{j,\al}$ on each site $j$ as
\begin{subequations}
\beqa
	a_{j,\al} &=& -i(c_{j,\al}-c^{\dg}_{j,\al}) \\
	b_{j,\al} &=& c_{j,\al}+c^{\dg}_{j,\al}\ .
\eeqa
\end{subequations}
In terms of these operators we have the identity
\beq
c^{\dg}_{j,\al} c_{j+1,\al} + c_{j,\al} c_{j+1,\al} + \text{h.c.}= i b_{j,\al}a_{j+1,\al}\ ,
\eeq
which holds for all $j\in\{1,\dots,L-1\}$. For all $j$, we also have the identity
\beq
	n_{j,\al}-\frac{1}{2} = -\frac{1}{2} i a_{j,\al}b_{j,\al}\ .
\eeq

These two identities are sufficient to completely rewrite $H_{\text{F}}$ in terms of the Majorana 
operators. For example, we have (at $t_{\al}=\Delta_{\al}$)
\beq
	H_{\text{MF}}(0)= -\sum_{\al=1}^2\frac{\Delta_{\al}}{2}\sum_{j=1}^{L-1}i b_{j,\al}a_{j+1,\al} + \sum_{\al=1}^2 \frac{\mu_{\al}}{2}\sum_{j=1}^L i a_{j,\al}b_{j,\al}\ .
\eeq
In addition, the number operator $\mathcal{N}_{w,\al}$ for wire $\al$ takes the form
\beq
	\mathcal{N}_{w,\al}= \frac{L}{2}-\frac{1}{2}\sum_{j=1}^L i a_{j,\al}b_{j,\al}\ ,
\eeq
and this identity can be used to rewrite the charging energy term in $H_{\text{F}}$ in terms of the 
Majorana operators.

We now turn to the JW transformation itself. 
For this transformation we introduce two sets of Pauli matrices 
$\left\{\sigma^x_{j,\al},\sigma^y_{j,\al},\sigma^z_{j,\al}\right\}$ (one 
set for each wire).
For the fermions on the first wire the transformation between the Majorana operators and the
Pauli matrices $\sigma^{x,y,z}_{j,\al}$ is
\begin{subequations}
\beqa
	a_{j,1} &=& \left(\prod_{k<j}\sigma^x_{k,1} \right)\sigma^z_{j,1} \\
	b_{j,1} &=& \left(\prod_{k<j}\sigma^x_{k,1} \right)\sigma^y_{j,1} \ .
\eeqa
\end{subequations}
For the second wire the relation is slightly more complicated because we have
to ensure that Majorana operators on two different wires still anticommute with each other.
Therefore, for the fermions on the second wire we have the transformation
\beqa
	a_{j,2} &=& \left(\prod_{\ell=1}^L\sigma^x_{\ell,1}\right)\left(\prod_{k<j}\sigma^x_{k,2} \right)\sigma^z_{j,2} \\
	b_{j,2} &=& \left(\prod_{\ell=1}^L\sigma^x_{\ell,1}\right)\left(\prod_{k<j}\sigma^x_{k,2} \right)\sigma^y_{j,2} \ .
\eeqa
To rewrite the fermionic Hamiltonian $H_{\text{F}}$ in terms of the spin operators, it will also be 
convenient to introduce the following notation for sums of $\sigma^x_{j,\al}$ operators: 
$\Sigma^x_{\al} = \sum_{j=1}^L\sigma^x_{j,\al}$. 

After the JW transformation, the mean-field part of the Hamiltonian takes the form
\beq
	H_{\text{MF}}(0)= -\sum_{\al=1}^2\frac{\Delta_{\al}}{2}\sum_{j=1}^{L-1}\sigma^z_{j,\al}\sigma^z_{j+1,\al} + \sum_{\al=1}^2 \frac{\mu_{\al}}{2}\Sigma^x_{\al}\ ,
\eeq
and the charging energy term is now
\begin{align}
	V &= \frac{1}{2}\sum_{\al=1}^2 \frac{\mathcal{E}_{c,\al}}{L}\left(\frac{1}{2}\Sigma^x_{\al} + N_{\al,0} - \frac{L}{2}\right)^2 \nnb \\
	 &+ \frac{\mathcal{E}_c'}{L}\left(\frac{1}{2}\Sigma^x_1 + N_{1,0} - \frac{L}{2}\right)\left(\frac{1}{2}\Sigma^x_2 + N_{2,0} - \frac{L}{2}\right)\ .
\end{align}
The entire Hamiltonian $H_{\text{F}}$ can then be rewritten as 
\begin{align}
	H_{\text{F}} &= \sum_{\al=1}^2\frac{\Delta_{\al}}{2}\sum_{j=1}^{L-1}\left(1 - \sigma^z_{j,\al}\sigma^z_{j+1,\al}\right) + \sum_{\al=1}^2 \frac{\td{\mu}_{\al}}{2}\Sigma^x_{\al} \nnb \\
	&+ \sum_{\al=1}^2\frac{\mathcal{E}_{c,\al}}{8L}\sum_{j,j'=1}^L\sigma^x_{j,\al}\sigma^x_{j',\al}(1-\delta_{j,j'}) \nnb \\
	&+ \frac{\mathcal{E}_c'}{4L}\Sigma^x_1 \Sigma^x_2 + \text{constant}\ , \label{eq:final-spin-Ham}
\end{align}
where the $\td{\mu}_{\al}$ are the modified chemical potentials that we defined in Eqs.~\eqref{eq:mu-mod}.
In particular, the form of the JW-transformed Hamiltonian explains why it is
$\td{\mu}_{\al}$, and not $\mu_{\al}$, that appears in our main result.

Finally, we need to discuss the various fermion parity operators and their form in
terms of spin operators. First, the total fermion parity operator $(-1)^{\mathcal{N}_w}$ takes the 
form\footnote{There is a minus sign error in Eq.~(4.3) of the Supplemental Material of Ref.~\onlinecite{LL}
that caused us to include an additional factor of $(-1)^L$ in the relation between the fermion parity
operator and the Ising symmetry operator. That sign error did not affect our results in
Ref.~\onlinecite{LL}, but we note here that the correct versions of Eq.~(4.3) and (4.5) of the Supplemental 
Material are $\hat{N}_w = \frac{L}{2} - \frac{1}{2}\sum_{j=1}^L\hat{\sigma}^x_j$ and $\hat{S} = (-1)^{\hat{N}_w}$, respectively (using the notation of Ref.~\onlinecite{LL}).}
\beq
	(-1)^{\mathcal{N}_w}= \mathcal{S}\ ,
\eeq
where $\mathcal{S}$ is the total Ising symmetry operator for both wires: 
$\mathcal{S}= \prod_{\al=1}^2\prod_{j=1}^L \sigma^x_{j,\al}$.
Next, we consider the fermion parity operators for each individual wire. Using the
JW transformation, we find that $\mathcal{N}_{w,\al}= \frac{L}{2}-\frac{1}{2}\Sigma^x_{\al}$, and 
so 
\beq
	\mathcal{P}_{\al} := (-1)^{\mathcal{N}_{w,\al}}= \mathcal{S}_{\al}\ ,
\eeq
where $\mathcal{S}_{\al}= \prod_{j=1}^L \sigma^x_{j,\al}$ is the Ising symmetry operator for wire $\al$.

If we put all of our results together, then we find that studying our original model in the subsector
of the $N$-particle sector where $\mathcal{P}_1 = \pm 1$ is equivalent to studying the
Hamiltonian $H_{\text{F}}$ (expressed in terms of spin operators) in the 
$\mathcal{S}_1 = \pm 1$ subsector of the sector with $\mathcal{S} = (-1)^N$.

\subsection{Completing the proof}

At this point we have mapped our model to a spin Hamiltonian with long-range 
$\sigma^x_{j,\al}\sigma^x_{j',\beta}$ interactions. We now show that this spin Hamiltonian is 
subject to the general stability result that we proved in Ref.~\onlinecite{LL-stability}.

To apply our result from Ref.~\onlinecite{LL-stability}, notice that the final
spin Hamiltonian from Eq.~\eqref{eq:final-spin-Ham} (after dropping the constant part) can be written in the 
general form
\beq
	H_{\text{spin}} =  \sum_{\al=1}^2\frac{\Delta_{\al}}{2}\sum_{j=1}^{L-1}\left(1 - \sigma^z_{j,\al}\sigma^z_{j+1,\al}\right) + \lambda W \ , \label{eq:general-spin-Ham}
\eeq
where $\lambda$ is a constant with units of energy and the perturbation term $W$ takes the form
\begin{align}
	W = \sum_{\al=1}^2 h_{\al}\Sigma^x_{\al} + \frac{1}{2}\sum_{j,j'=1}^L\sum_{\al,\beta=1}^2 f^{(\al,\beta)}(|j-j'|)\sigma^x_{j,\al}\sigma^x_{j',\beta}\ . \label{eq:general-interaction}
\end{align}
Here, $h_{\al}$ and $f^{(\al,\beta)}(|j-j'|)$ are dimensionless coefficients, and we assume that
$f^{(\al,\beta)}(0) = 0$ if $\al = \beta$ to avoid including constant terms in $W$. In particular, we can choose the
following values for $\lambda$, $h_{\al}$, and $f^{(\al,\beta)}(|j-j'|)$:
\begin{subequations}
\begin{align}
\lambda = \max(|\mathcal{E}_{c,1}|, &|\mathcal{E}_{c,2}|, |\mathcal{E}_c'|, |\td{\mu}_1|, |\td{\mu}_2|)  \\
h_{\al} &= \frac{\td{\mu}_{\al}}{2\lambda } \\ 
f^{(1,1)}(|j-j'|) &= \frac{\mathcal{E}_{c,1}}{4\lambda L}(1-\delta_{j,j'})\\
f^{(2,2)}(|j-j'|) &= \frac{\mathcal{E}_{c,2}}{4\lambda L}(1-\delta_{j,j'}) \\
f^{(1,2)}(|j-j'|) &= \frac{\mathcal{E}_{c}'}{4\lambda L} \\
f^{(2,1)}(|j-j'|) &= f^{(1,2)}(|j-j'|)\ . 
\end{align}
\end{subequations}

Notice that the coefficients $f^{(\al,\beta)}(|j-j'|)$ are bounded in magnitude by $\frac{1}{4L}$ and therefore, in particular,
\beq
	|f^{(\al,\beta)}(|j-j'|)|\leq 1\ . \label{eq:bounded-coeffs}
\eeq
Also, for any site $j$ the coefficients $f^{(\al,\beta)}(|j-j'|)$ satisfy a \emph{summability condition} of the form
\beq
 \sum_{j',\al,\beta}\left|f^{(\al,\beta)}(|j-j'|)\right| \leq c \ , \label{eq:summability}
\eeq
where $c$ is a constant that can be chosen to be independent of $L$. (Specifically, in this case, we can take $c = 1$). 

With this setup, we are now in a position to apply the stability result from Ref.~\onlinecite{LL-stability} (adapted to the current setting).

\begin{thm*}[Adapted from Theorem 1 of Ref.~\onlinecite{LL-stability}]
\label{thm:stability-LL}
Consider a Hamiltonian $H_{\text{spin}}$ of the form \eqref{eq:general-spin-Ham} with interaction coefficients that satisfy \eqref{eq:bounded-coeffs} and 
\eqref{eq:summability}. There exists a $L$-independent constant 
$\lambda_0 > 0$ such that, if $|\lambda| < \lambda_0$, then (1) $H_{\text{spin}}$
has a unique ground state and a finite energy gap in each subsector of the $\mathcal{S} = (-1)^N$ sector
with fixed $\mathcal{S}_1$ eigenvalue, and (2) the ground
state energy splitting $|E_{+}(\lambda)-E_{-}(\lambda)|$ between the $\mathcal{S}_1 = \pm 1$ subsectors 
of the $\mathcal{S} = (-1)^N$ sector satisfies the exponential bound
\beq
	|E_{+}(\lambda)-E_{-}(\lambda)| \leq c_1 L e^{-c_2 L}\ , 
\eeq
where $c_1$ and $c_2$ are positive constants that depend on $\Delta_{\al}$, $h_{\al}$, and $\lambda$, 
but not on $L$. 
\end{thm*} 

Combining the above theorem with the mappings discussed in the previous sections, Theorem~\ref{thm:main} follows immediately.

A final comment: as we mentioned earlier, the above theorem is similar, but not identical, to Theorem 1 of Ref.~\onlinecite{LL-stability}. The main difference is that the original theorem was phrased in terms of a single spin chain, while the above theorem involves \emph{two} coupled spin chains. Also, while the original theorem assumed periodic boundary conditions, here we consider a model with \emph{open} boundary conditions. These minor modifications in the statement of the theorem can be accommodated by making minor technical changes to the proofs in Ref.~\onlinecite{LL-stability}. For completeness we summarize these changes in Appendix~\ref{app:changes-to-proof}. 

\section{Conclusion}
\label{sec:discussion-conclusion}

In this paper we have completed the work that we began in Ref.~\onlinecite{LL}: combining the results
of this paper and Ref.~\onlinecite{LL}, we have now proved that three of the key properties of 
the mean-field TSC models also hold in a number-conserving toy model of a qubit formed from a 1D 
TSC. These properties are: (i) the existence of two low-energy ``qubit'' states that 
are separated from the rest of the spectrum by a finite energy gap; (ii) an exponentially small bound on the energy splitting of the qubit states (the main contribution of this paper); and (iii) the existence of long-range ``Majorana-like'' correlations between the ends of the fermionic wires in the 1D
TSC model. These results provide an important proof of principle that the key properties of the mean-field TSC
models --- properties that are important for quantum computing applications --- can indeed be realized in a 
number-conserving model without fine-tuning.

A natural next step would be to study the above properties in a more realistic model. Such a model would include some or all of the following ingredients: first, the bulk s-wave SC would be modeled by a 3D number-conserving Hamiltonian of interacting fermions (rather than a single number and phase degree of freedom). Second, the 1D wires would couple to the 3D bulk via single-particle hopping terms (rather than pair hopping terms). Finally, the fermions would interact via microscopic Coulomb interactions (rather than an effective charging energy term). While it is not obvious how to generalize our techniques to this setting, we see this as an interesting direction for future work.

\acknowledgements

M.F.L. and M.L. acknowledge the support of the Kadanoff Center for Theoretical Physics at the University of 
Chicago. This work was supported in part by a Simons Investigator grant (M.L.) and by the Simons Collaboration on Ultra-Quantum Matter, which is a grant from 
the Simons Foundation (651440, M.L.).

\appendix

\section{Applying our stability results to the two-wire model in this paper}
\label{app:changes-to-proof}

As we mentioned in the main text, we need to make a few changes to our setup and proofs from
Ref.~\onlinecite{LL-stability} in order to apply the stability results from that paper to the
spin model from Eq.~\eqref{eq:final-spin-Ham}. There are two main 
differences between this spin model and the model that we studied in Ref.~\onlinecite{LL-stability} 
(see Eqs.~(2.1)-(2.3) of that paper). The first difference is that the unperturbed spin model in this paper 
consists of \emph{two} transverse-field Ising chains, instead of one, and the long-range interaction
term (which comes from the charging energy term) couples spins within \emph{and} between the two chains.
The second difference is that the model in this paper has open boundary conditions for the two chains,
while we assumed periodic boundary conditions in Ref.~\onlinecite{LL-stability}.

We now give a brief list of the main changes that we need to make in the setup and proofs
of Ref.~\onlinecite{LL-stability} in order to apply the results of that paper to 
Eq.~\eqref{eq:final-spin-Ham} of this paper. In what follows we assume that any reader of this 
appendix is familiar with the concepts and notation from Ref.~\onlinecite{LL-stability}.

Changes needed because of the open boundary conditions:
\begin{itemize}
\item Sums over the separation $r = |j-j'|$ between sites $j$ and $j'$ on the spin chains
now run from $1$ to $L-1$ instead of $1$ to $L/2$.
\item The size of the blocked spacetime lattice $\Lambda$ used in the polymer expansion is now $|\Lambda| = (2L - 1)M$ instead of $2LM$. 
Compared to the periodic case, the lattice is now missing one column of plaquettes.
\item Support sets can now include ``half-boxes'' at the boundaries of the system. These
do not introduce any complications and they can be treated exactly the same as the full boxes
in the bulk of the system.
\end{itemize}

Changes needed because we now have two wires (two spin chains):
\begin{itemize}
\item The weights $W_{\pm}(X)$ are now for the system restricted to the $\mathcal{S}_1 = \pm 1$
subsectors of the $\mathcal{S} = (-1)^N$ sector of the Hilbert space for the two spin chains.
\item We now define the macroscopic support set $s(\mathcal{C})$ of a microscopic domain wall and
interaction configuration $\mathcal{C}$ as follows.
\begin{enumerate}
\item If $\mathcal{C}$ has a domain wall worldline (from either wire) that passes all the way through a
given plaquette, then that plaquette is included in $s(\mathcal{C})$.
\item If a perturbation term $V_Y$ has support on site $j$ (on either wire) within the time interval
$[\tau(\ell-1),\tau\ell]$, then $s(\mathcal{C})$ contains the box or half-box centered on $j$ within
that time slice.
\item If a perturbation term $V_Y$ has support on site $j$ (on either wire) \emph{and} site $k$ 
(on either wire) within the time interval $[\tau(\ell-1),\tau\ell]$, then $s(\mathcal{C})$ contains a 
\emph{single} dashed line connecting the boxes or half-boxes centered on sites $j$ and $k$ within
that time slice. Of course, this rule is only relevant for the $\sigma^x\sigma^x$ terms.
\end{enumerate}
\item With the new definition of $s(\mathcal{C})$ from the previous bullet point, there are now more 
microscopic configurations that lead to the same macroscopic support set, and we need to be sure to sum over 
these additional configurations when deriving the bound on the weights 
$W_{\pm}(X)$ from Lemma 1 of Ref.~\onlinecite{LL-stability}. In particular, this means that
the ``partial perturbation terms'' $\mathcal{V}_{\ell}$ from Eq.~(A17) of Ref.~\onlinecite{LL-stability} 
now contain more terms, and the constant $K$ that first appears in Eq.~(A23) will be larger. 
\item The parameter $\Delta$ that appears in the upper bound in Eq. (A4) of 
Ref.~\onlinecite{LL-stability} should be chosen as $\Delta = \min(\Delta_1,\Delta_2)$. The reason
is that this upper bound is now obtained by taking the smaller of the two excitation energies for domain 
walls in the unperturbed Hamiltonian. 
\item The parameter $|h|$ in Eq. (A7) of Ref.~\onlinecite{LL-stability} should be chosen as
$|h| = \max(|h_1|,|h_2|)$.
\item In Eq. (A7) of Ref.~\onlinecite{LL-stability}, the weights in the ``$+$'' and
``$-$'' Ising sectors are both bounded from above by an expression that features a trace over only the 
``$+$'' Ising sector. Instead of this, we now have a bound in which the weights in the $\mathcal{S} = (-1)^N$,
$\mathcal{S}_1 = \pm 1$ subsectors are both bounded from above by an expression that
features a trace in the $\mathcal{S} = 1$, $\mathcal{S}_1 = 1$ subsector.
\end{itemize}


%

\end{document}